\documentclass[12pt,letterpaper]{article}
\pdfoutput=1
\usepackage[utf8]{inputenc}
\usepackage{amsmath}
\usepackage{slashed}
\usepackage{amssymb}
\usepackage{amsthm}
\usepackage{xcolor}
\definecolor{nicered}{rgb}{0.7,0.1,0.1}
\definecolor{nicegreen}{rgb}{0.1,0.5,0.1}
\usepackage[colorlinks=true,citecolor= nicegreen,linkcolor=nicered]{hyperref}
\usepackage{graphicx}
\graphicspath{{figures/}}
\usepackage{siunitx} 
\usepackage{cancel}

\usepackage{comment}
\includecomment{details}
\specialcomment{details}
{\begingroup}{\endgroup}

\newcommand{\DetailsWeylconventions}{}
\newcommand{\DetailsWeylspinors}{}
\newcommand{\DetailsRddoublet}{}
\newcommand{\DetailsRudoublet}{}
\newcommand{\Detailscontraction}{}
\newcommand{\DetailsInteractionBasis}{}

\newcommand{\DetailsDiracAndMajoranaFermions}{}
\newcommand{\DetailsNeutralWyelLagrangian}{}
\newcommand{\DetailsNeutrinoMassInteractionBasis}{}
\newcommand{\DetailsNeutrinoMassEigenBasis}{}
\newcommand{\DetailsPassarinoVeltman}{}
\newcommand{\DetailsMutoEGamma}{}

\excludecomment{details}

\title{\vspace*{-6em}
  \begin{flushright}
    {\sf\small TAUP2993/15}
  \end{flushright}
\vspace*{2em}
Radiative neutrino masses in the singlet-doublet fermion dark
  matter model with scalar singlets}

\author{Diego Restrepo\footnote{\href{mailto:restrepo@udea.edu.co}{restrepo@udea.edu.co}},
  Andrés Rivera\footnote{\href{mailto:afelipe.rivera@udea.edu.co}{afelipe.rivera@udea.edu.co}}, Marta Sánchez-Peláez\footnote{\href{mailto:martal.sanchez@udea.edu.co}{martal.sanchez@udea.edu.co}},  Oscar Zapata\footnote{\href{mailto:oalberto.zapata@udea.edu.co}{oalberto.zapata@udea.edu.co}}\\
\textit{\small  Instituto de F\'{i}sica, Universidad de Antioquia,} \\
\textit{\small  Calle 70 No. 52-21, Medell\'{i}n, Colombia}\\[4mm]
Walter Tangarife\footnote{\href{mailto:waltert@post.tau.ac.il }{waltert@post.tau.ac.il }}\\
\textit{\small Department of Particle Physics, School of Physics and Astronomy, }\\
\textit{\small Tel Aviv University, Tel Aviv, 69978, Israel}
}
\date{\small April 28, 2015}
\begin{document}

\maketitle
\begin{abstract}
  When the singlet-doublet fermion dark matter model is extended with
  additional $Z_2$--odd real  singlet scalars, neutrino masses and mixings
  can be generated at one-loop level.  In this work, we discuss the salient 
 features arising from the combination of the two resulting
  simplified dark matter models.  When the $Z_2$-lightest odd particle is a
  scalar singlet, $\operatorname{Br}(\mu\to e \gamma)$ could be
  measurable provided that the singlet-doublet fermion mixing is small
  enough. In this scenario, also the new decay channels of vector-like
  fermions into scalars can generate interesting leptonic plus
  missing transverse energy signals at the LHC. On the other hand, in
  the case of doublet-like fermion dark matter, scalar coannihilations
  lead to an increase in the relic density which allow to lower the bound of
  doublet-like fermion dark matter.
\end{abstract}
\section{Introduction}

In view of the lack of signals of new physics in strong production at the
LHC, there is a growing interest in simplified models where the
production of new particles is only through electroweak processes,
with lesser constraints from LHC limits.
In particular, there are simple standard model (SM) extensions with
dark matter (DM) candidates, such as the singlet scalar dark
matter (SSDM)
model~\cite{Silveira:1985rk,McDonald:1993ex,Burgess:2000yq}, or the
singlet-doublet fermion dark matter (SDFDM)
model~\cite{ArkaniHamed:2005yv,Mahbubani:2005pt,D'Eramo:2007ga,Enberg:2007rp,Cohen:2011ec,Cheung:2013dua}.
In this kind of models, the prospects for signals at LHC are in general
limited because of the softness of final SM particles coming from the
small charged to neutral mass gaps of the new particles, which is usually required 
to obtain the proper relic density.
In this sense, the addition of new particles, motivated for example by
neutrino physics, could open new detection possibilities, either
trough new decay channels or additional mixings which increase the
mass gaps.

On those lines, scotogenic models~\cite{Ma:2006km}, featuring neutrino masses suppressed
by the same mechanism that stabilizes dark matter, are being
thoroughly studied with specific predictions in almost all the
current terrestrial and satellite detector experiments (For a review
see for example~\cite{Boucenna:2014zba}).
The simplest models correspond to extensions of the inert doublet
model~\cite{Deshpande:1977rw,Barbieri:2006dq} with extra singlet or
triplet fermions.
Recently, the full list of 35 scotogenic models with neutrino masses at
one-loop~\cite{Ma:1998dn,Bonnet:2012kz}\footnote{The general
  realization of the Weinberg operator at two-loops have been undertaken
  in~\cite{Sierra:2014rxa}}, and at most triplet representations of
$SU(2)_L$, was presented in~\cite{Restrepo:2013aga} (and partially
in~\cite{Law:2013saa}).
The next to simplest scotogenic model is possibly the one where the
role of the singlet fermions is played by singlet scalars, and the role
of the scalar inert doublet is played by a vector-like doublet fermion.
One additional singlet fermion is required to generate neutrino masses
at one-loop level.
This kind of extension of the singlet dark matter model is labeled
as the model T13A with $\alpha=0$ in \cite{Restrepo:2013aga}. The
extra fermion, required in order to have radiative neutrino masses, can
be the singlet in the SDFDM model.

In the simplest scotogenic model~\cite{Ma:2006km}, singlet fermion
dark matter is possible but quite restricted by lepton flavor
violation (LFV)~\cite{Toma:2013zsa,Vicente:2014wga}. 
In contrast, we will show that in the present model the region of the
parameter space, corresponding to fermion dark matter, is well below
the present and near future constraints on $\operatorname{Br}(\mu\to
e\gamma)$.

On the other hand, when the lightest $Z_2$-odd particle (LOP) is one
of the scalar singlets, in the regions of the parameter space 
compatible with constraints from LFV, we could have promising signals at colliders,
thanks to the electroweak production of fermion doublets and possible
large branchings into charged leptons.

The dark matter phenomenology of both the SSDM and SDFDM models has
been extensively studied in the literature and recently
revisited in~\cite{Abe:2014gua}.
Here we consider the possible effect of coannihilations with the
scalar singlets for fermion dark matter. 
We will see that these coannihilations tend to increase the relic
density of dark matter and may modify the viable parameter space of
the model. 
Specifically, they allow to reduce the lower bound on the mass of the
doublet-like dark matter particle from around $\SI{1100}{GeV}$ down to
about $\SI{900}{GeV}$.

The rest of the paper is organized as follows.  In the next section,
we present the model.  Our main results are presented in Sections
\ref{sec:one-loop-neutrino} to \ref{sec:singlet-doublet-dark} where we
describe the correlation between the generation of neutrino masses and
lepton flavor violation, new signals at colliders in the case of
scalar dark matter, and new coannihilation possibilities in the case
of singlet-doublet fermion dark matter. Finally, in section
\ref{sec:conclusions} we present our conclusions.  In the Appendix we
present the analytic diagonalization formulae for the mass matrix of
neutral fermions.

\section{The model}
\label{sec:model}
The particle content of the model consists of two $SU(2)_L$-doublets of
Weyl fermions $\widetilde{R}_u$, $R_d$  with opposite hypercharges; one singlet
Weyl fermion $N$ of zero hypercharge, and a set of 
real scalar singlets  $S_{\alpha}$ also of zero hypercharge. 
All of them are odd under one imposed $Z_2$ symmetry, under which the SM
particles are even. 
The new particle content is summarized in Table~\ref{tab:partcont}.
\begin{table}
  \centering
  \begin{tabular}{|l|l|l|l|}
    \hline  
    Symbol     & $\left( SU(2)_L, U(1)_Y \right)$ & $Z_2$ & \text{Spin}\\ \hline
    $S_{\alpha}$ & $(1,0)$ & $-$ & 0\\
    $N$  & $(1,0)$ & $-$ & 1/2\\
     $\widetilde{R}_u$, & $(2, +1/2)$ & $-$ & 1/2\\ 
     $R_d$ & $(2, -1/2)$ & $-$ & 1/2\\ \hline
  \end{tabular}
  \caption{$\alpha$-set of scalars and Weyl fermions of the model.}
  \label{tab:partcont}
\end{table}
The most general $Z_2$-invariant Lagrangian is given by
\begin{align}
\label{eq:lt13a}
 \mathcal{L}= &\mathcal{L}_{\text{SM}}+ M_D \epsilon_{ab}R^a_d \widetilde{R}^b_u-\tfrac{1}{2}M_N NN-h_{i\alpha} \epsilon_{ab}\widetilde{R}_u^a L_{i}^b S_{\alpha}-\lambda_d\, \epsilon_{ab}H^a R_d^b N-\lambda_u \epsilon_{ab}\widetilde{H}^a \widetilde{R}_u^b N+\text{h.c}\nonumber\\
&-\left[ 
\tfrac{1}{2}\left({M}_S^2\right)_{\alpha\beta} S_{\alpha}S_\beta
   +\lambda^{SH}_{\alpha\beta} \epsilon_{ab}\widetilde{H}^{a}H^bS_{\alpha}S_{\beta}+\lambda^{S}_{\alpha\beta\gamma\delta}S_{\alpha}S_{\beta}S_{\gamma}S_{\delta} 
\right], 
\end{align}
where $L_{i}$ are the lepton doublets,  we have defined the new $SU(2)_L$--doublets in terms of left-handed Weyl fermions as
\begin{align*}
  R_{d}=&
\DetailsRddoublet
  \begin{pmatrix}
    \psi_{L}^{0}\\
    \psi_{L}^{-}
  \end{pmatrix}
&  
\widetilde{R}_{u}=&
\DetailsRudoublet
  \begin{pmatrix}
   - \left( \psi_{R}^{-} \right)^{\dagger}\\
     \left(\psi_{R}^{0}\right)^{\dagger}
  \end{pmatrix},
\end{align*}
and $H=\begin{pmatrix}0 & (h+v)/\sqrt{2}\end{pmatrix}^{\operatorname{T}}$ as the SM Higgs doublet with $\widetilde{H}=i\sigma_2H^*$ and $v=\SI{246}{GeV}$.
\DetailsWeylconventions
\DetailsWeylspinors
\Detailscontraction 
\DetailsInteractionBasis
In the scalar potential, we assume that the $\mathbf{M}_{S}^{2}$ matrix has only
positive entries and $\left(M_S^2 \right)_{\alpha\beta}+\lambda^{SH}_{\alpha\beta}v^2=0$ for $\alpha\ne\beta$,  which means $S_\alpha$ are mass eigenstates with masses $m_{S_{\alpha}}^{2}=\left({M}_S^2 \right)_{\alpha\alpha}+\lambda^{SH}_{\alpha\alpha}v^2$ and $m_{S_\alpha}<m_{S_{\alpha+1}}$.
\DetailsDiracAndMajoranaFermions
\begin{details}
\subsection{Mass matrices}
We return back to the Lagrangian in terms of Weyl spinors of eq.~\eqref{eq:lt13a}.  
\end{details}
On the other hand, the $Z_2$-odd fermion spectrum is composed by a
charged Dirac fermion $\chi^-=(\psi^-_L,\, \psi^-_R)^T$ with a tree
level mass $m_{\chi^\pm}=M_D$, and three Majorana fermions arisen from
the mixture between the neutral parts of the $SU(2)_L$ doublets and
the singlet fermion. 
By defining the fermion basis through the vector
$\boldsymbol{\Xi}=\left(N,\psi_L^0,\left( \psi_R^0 \right)^{\dagger}
\right)^T$,
 \begin{details}
The neutral-fermion part of the Lagrangian is
(by using eqs.~\eqref{eq:mdpsipsi}, \eqref{eq:HRdN}, \eqref{eq:HRuN}) %see details.tex
\begin{align*}
  -\mathcal{L}_{\Xi}=&\tfrac{1}{2}M_N NN+M_D{\psi_{R}^0 }^{\dagger}\psi_{L}^0+
h_{i\alpha}{\psi_{R}^0}^{\dagger}\nu_{Li}S_{\alpha}+\lambda_d  N H^{0}\psi_L^{0}
+\lambda_u \psi^{0\dagger}_R H^{0*} N+\text{h.c} \nonumber\\
\supset& \frac{1}{2}M_N NN+M_D{\psi_{R}^0 }^{\dagger}\psi_{L}^0
+h_{i\alpha}{\psi_{R}^0}^{\dagger}\nu_{Li}S_{\alpha}
+\frac{\lambda_d v}{2\sqrt{2}} \left(\psi_L^{0} N +N\psi_L^{0}\right)
+\frac{\lambda_u v}{2\sqrt{2}}\left(\psi^{0\dagger}_R N+N\psi^{0\dagger}_R  \right) 
+\text{h.c}\,,
\end{align*}
\end{details}
the neutral fermion mass matrix reads
\begin{align}
\label{eq:Mchi}
  \mathbf{M}^{\chi}=\begin{pmatrix}
 M_N                 &-m_{\lambda}\cos\beta&m_{\lambda}\sin\beta\\
-m_{\lambda}\cos\beta &  0                  & -M_D\\
m_{\lambda}\sin\beta&  -M_D                &  0  \\
\end{pmatrix},
\end{align}
where
\begin{align}
  \label{eq:etabeta}
m_{\lambda}=&  \frac{\lambda v }{\sqrt{2}}\,,&
  \lambda=&\sqrt{\lambda_u^2+\lambda_d^2}\,,&
  \tan\beta=&\frac{\lambda_u}{\lambda_d}\,.
\end{align}
The specific signs on the right hand side of eq.~\eqref{eq:lt13a} were chosen so
that the terms in the mass matrix $ \mathbf{M}^{\chi}$ follow the same conventions as in the
neutralino mass matrix~\cite{Martin:2012us}.  
From this follows that the supersymmetric case corresponds to the pure bino-higgsino limit
with $m_\lambda=m_Z\sin\theta_W$ leading to $\lambda=g'/\sqrt{2}$. 
The Majorana fermion mass eigenstates $ \mathbf{X}=(\chi_1,\chi_2,\chi_3)^T$ are obtained through the rotation matrix $\mathbf{N}$ as $\boldsymbol{\Xi}=\mathbf{N}\mathbf{X}$
such that
\begin{align}
\label{eq:chidiag}
 \mathbf{N}^{\operatorname{T}}\mathbf{M}^\chi \mathbf{N}=\mathbf{M}^\chi_\text{diag},
\end{align}
%The usual suppersymmetric convention is E=N'^{\dagger}X -> N'^{*}M^{\chi}N'^{\dagger}=M^{\chi}_{diag}.
with
$\textbf{M}^\chi_{\text{diag}}=\operatorname{Diag}(m^\chi_1,m^\chi_2,m^\chi_3)$ and $m^\chi_n$ being the corresponding masses (no mass ordering is implied). 
In which follows we assume CP invariance and therefore $\mathbf{N}$ can be chosen real.
The analytical diagonalization of the neutral fermion mass matrix is
carried out in Appendix~\ref{sec:analyt-form-mass}. For the subsequent
analysis, it will be convenient to have some approximate expressions in
the limit of small doublet-fermion mixing ($m_\lambda\ll M_D,M_N)$.
Expanding the analytical expressions for the eigensystem of
eq.~\eqref{eq:chidiag} given in Appendix~\ref{sec:analyt-form-mass},
up to order $m_{\lambda}^2$, the fermion masses are
\begin{align}
\label{eq:ml2}
m^\chi_1=&M_{N} + \frac{M_{D} \sin{\left (2 \beta \right )} + M_{N}}{M_{N}^{2}- M_{D}^{2} }\, m_{\lambda}^{2}+\mathcal{O}\left( m_{\lambda}^4 \right) \nonumber\\
m^\chi_2=&M_{D} + \frac{ \sin(2 \beta ) + 1}{2 \left( M_{D} -  M_{N} \right)}\,m_{\lambda}^{2}+\mathcal{O}\left( m_{\lambda}^4 \right) \nonumber\\
m^\chi_3=&- M_{D} + \frac{ \sin(2 \beta ) - 1}{2 \left( M_{D} + M_{N} \right) }\,m_{\lambda}^{2}+\mathcal{O}\left( m_{\lambda}^4 \right)\,.
\end{align}
Approximate expressions for the mixing matrix are also given in that Appendix.
\begin{details}
The neutral Lagrangian in the mass-eigenstate basis includes

\DetailsNeutralWyelLagrangian

\begin{align}
\label{eq:masseig}
   \mathcal{L}_0=&h_{i\alpha}N_{3n}\chi_n \nu_{Li}S_\alpha+\frac{1}{2}m_{\chi_n}\chi_n\chi_n +\text{h.c}\,.
\end{align}
\end{details}

\section{One-loop neutrino masses}
\label{sec:one-loop-neutrino}
\begin{figure}
  \centering
  \includegraphics[scale=0.4]{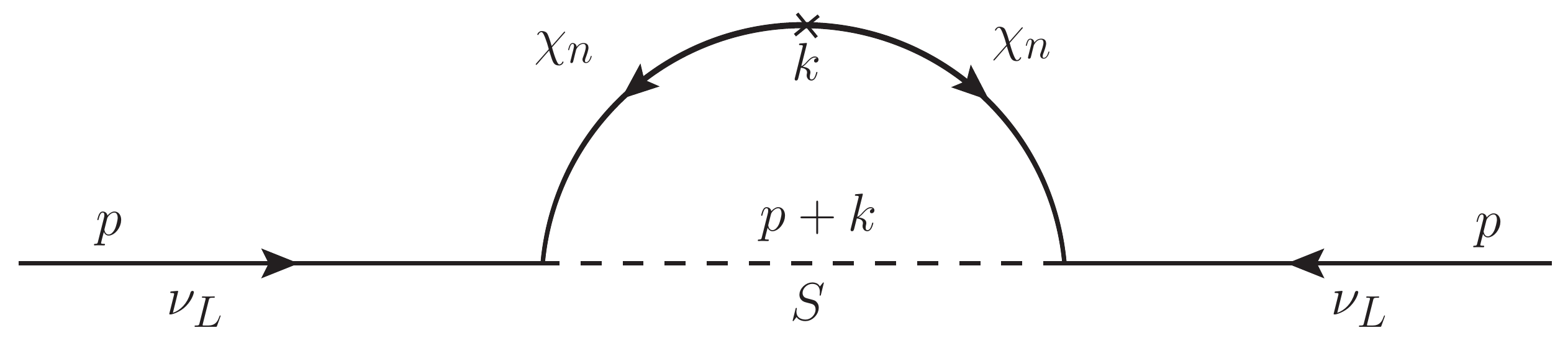} 
  \caption{One-loop Weyl-spinor Feynman rules~\cite{Dreiner:2008tw} for 
    the contributions to a neutrino mass, with three Majorana fermions $n=1,2,3$, and a singlet scalar $S$.
}
  \label{fig:T13Aweylme}
\end{figure}

By assigning a null lepton number to the new fields in the
model\footnote{If complex singlets instead real singlets are
  considered, an accidentally conserved lepton number would have been
  obtained in the
  Lagrangian, and such a case
  vanishing neutrino masses are expected.}, the only lepton-number
violating term in the Lagrangian eq. (\ref{eq:lt13a})  is the one with
coupling $h_{i\alpha}$. 
Hence, the introduction of real singlet scalars allows to generate
non-zero neutrino masses at one-loop level through the diagram shown
in figure~\ref{fig:T13Aweylme}. 
\begin{details}
By using the Feynman rules for Weyl spinors~\cite{Dreiner:2008tw}, we
have the diagrams for neutrino masses at one-loop shown in figure
\ref{fig:t13aweyl}.
\begin{figure}
  \centering
\includegraphics[scale=0.3]{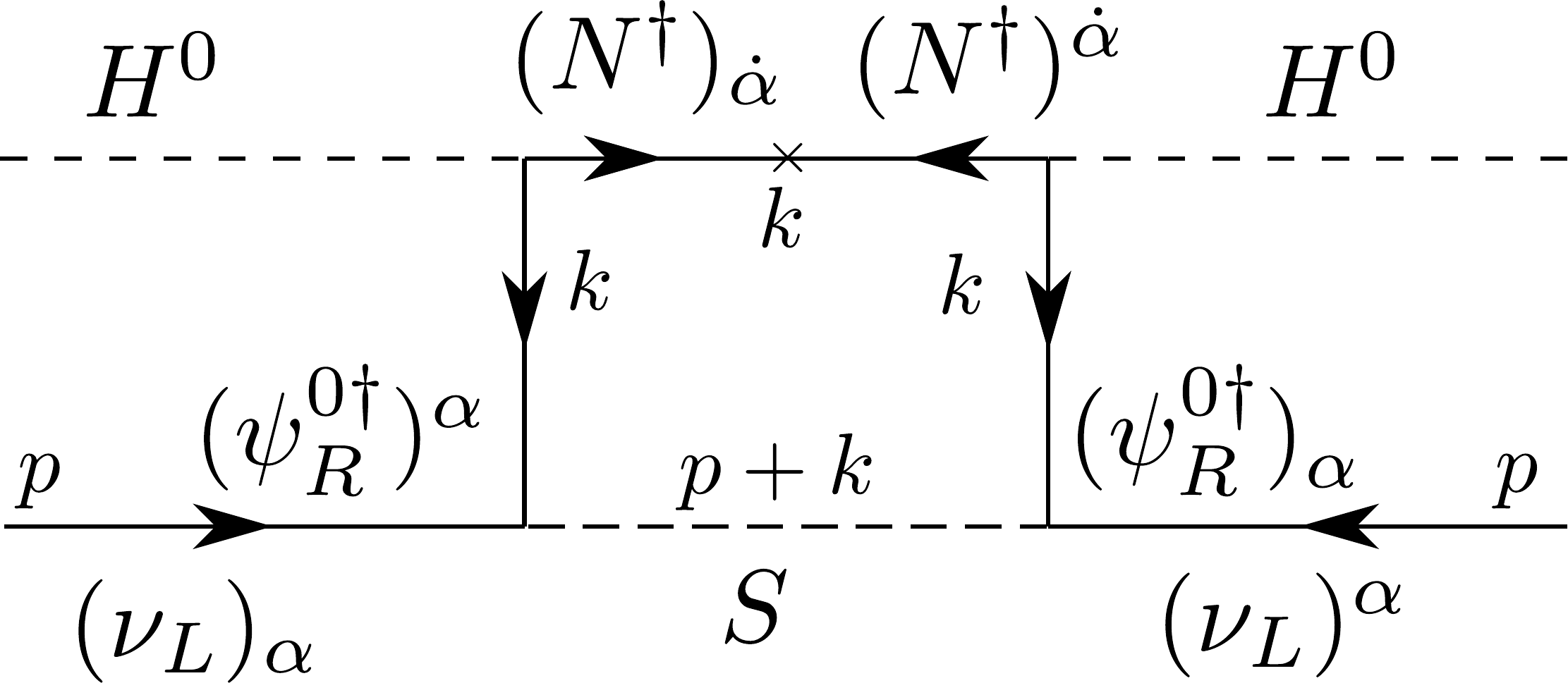}\qquad \includegraphics[scale=0.3]{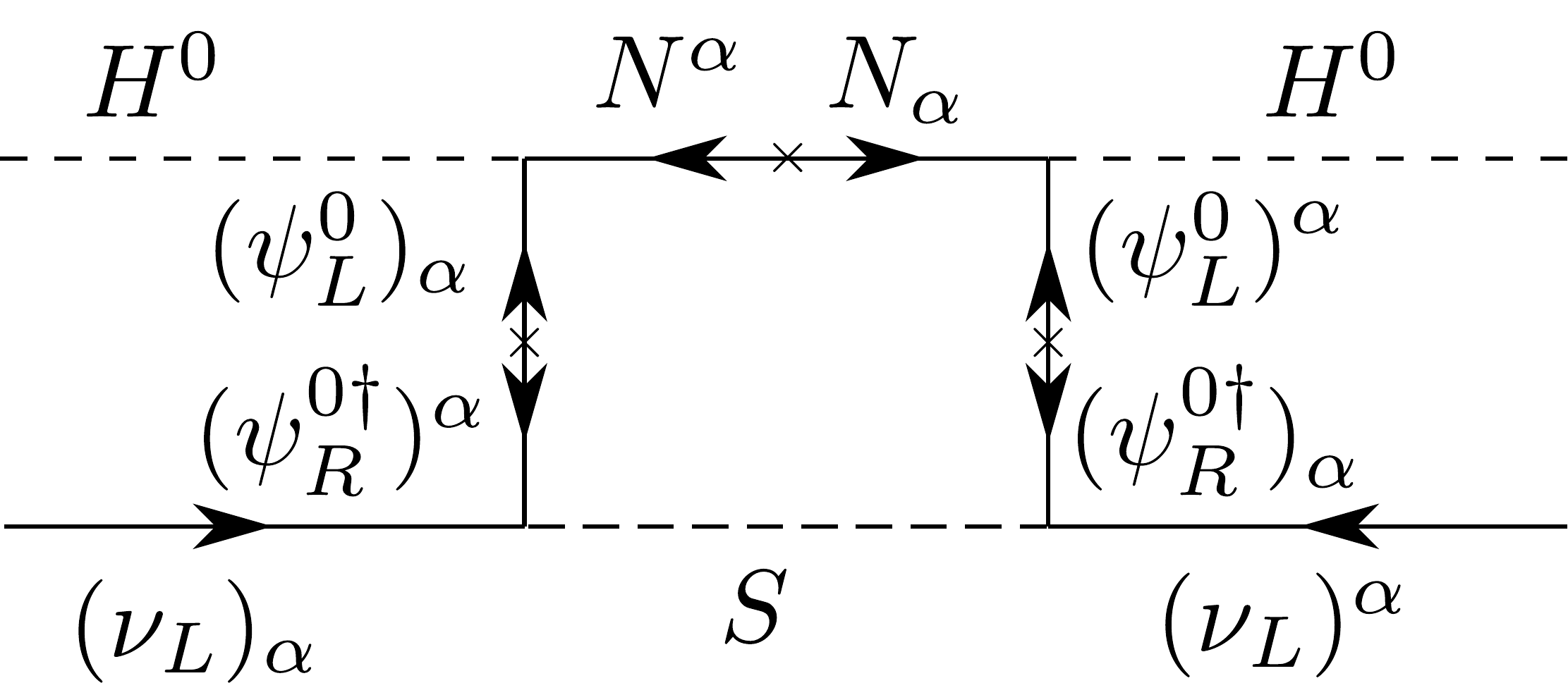}
  \caption{one-loop neutrino mass in the interaction basis.}
  \label{fig:t13aweyl}
\end{figure}
\end{details}
The resulting one-loop neutrino mass matrix was presented in the
interaction basis in \cite{Bonnet:2012kz} and \cite{Suematsu:2010nd},
and more recently in the limit $\lambda_{d}=0$ and $M_N\to 0$ in
\cite{Fraser:2014yha}.
\DetailsNeutrinoMassInteractionBasis
Instead, we work out the calculations in the more convenient mass-eigenstate basis, in which
the neutrino mass matrix takes the form 
\DetailsNeutrinoMassEigenBasis
\begin{align}
\label{eq:mnueig}
  M^{\nu}_{ij}=-&\sum_{\alpha}\frac{h_{i\alpha}h_{j\alpha}}{16\pi^2}\sum_{n=1}^3 \left( N_{3n} \right)^2m_{\chi_n}\; B_0 \left(0;m_{\chi_n}^2,m^2_{S_{\alpha}} \right),
\end{align}
where $B_0\left
(0;m_{\chi_n}^2,m^2_{S_{\alpha}} \right)$ is the $B_0$ Passarino-Veltman function~\cite{Passarino:1978jh} and $ (N_{mn})$ are matrix elements of the rotation matrix $\mathbf{N}$.
\DetailsPassarinoVeltman
By using  the identity 
\begin{align}
\label{eq:divcan}
  \sum_{n=1}^{3} \left( N_{3n} \right)^2m^\chi_n=\left( \mathbf{M}^\chi \right)_{33}=0,
\end{align}
we obtain the expected cancellation of divergent terms coming from the mass independent term in $B_0$, leading to the finite neutrino mass matrix
\begin{details}
  and \eqref{eq:mnueigdev}  %see details.tex
\end{details}
\begin{align}
   M^{\nu}_{ij}=&\sum_{\alpha}\frac{h_{i\alpha}h_{j\alpha}}{16\pi^2}\sum_{n=1}^3 \left( N_{3n} \right)^2m_{\chi_n}
\,f\left( m_{S_\alpha},m_{\chi_n} \right) ,\\\label{eq:Mnuij}
=&\sum_{\alpha} h_{i\alpha} \Lambda_{\alpha} h_{j\alpha}\\\label{eq:CI}
=&\left( \mathbf{h}\mathbf{\Lambda}\mathbf{h}^{\operatorname{T}} \right)_{ij}\,,
\end{align}
with
$f \left( m_1,m_2 \right)= (m_1^2\,\ln  m_1^2 -m_2^2\,\ln m_2^2 )/(m_1^2-m_2^2)$, $\boldsymbol{\Lambda}=\operatorname{Diag}\left(\Lambda_1,\Lambda_2,\Lambda_3\right)$ and
\begin{align}
\label{eq:Lambda}
  \Lambda_\alpha=&\frac{1}{16\pi^2}\sum_{n=1}^3 \left( N_{3n} \right)^2m_{\chi_n}\,f\left( m_{S_\alpha},m_{\chi_n} \right).
\end{align}
The flavor structure of the neutrino mass matrix $M^{\nu}_{ij}$, given
by eq. (\ref{eq:CI}), allows us to express the Yukawa couplings in
terms of the neutrino oscillation observables (ensuring the proper
compatibility with them) through the Casas-Ibarra parametrization introduced in
\cite{Casas:2001sr,Ibarra:2003up}. 
Thus, by using an arbitrary complex orthogonal rotation matrix
$\boldsymbol{\mathcal{R}}$, the Yukawa couplings $h_{i\alpha}$ are
given by
\begin{align*}
  \mathbf{h}^{T}=\mathbf{D}_{\sqrt{{\Lambda}^{-1}}}\,\boldsymbol{\mathcal{R}}\,\mathbf{D}_{\sqrt{m_{\nu}}}\,U^{\dagger} \,,
\end{align*}
where $\mathbf{D}_{\sqrt{m_\nu}}=\operatorname{Diag}
\left(\sqrt{m_{\nu 1}} , \sqrt{m_{\nu2}},\sqrt{m_{\nu3}}\right)$,
$\mathbf{D}_{\sqrt{\Lambda^{-1}}}=\operatorname{Diag}
\left(\sqrt{\Lambda_1^{-1}} , \sqrt{\Lambda_2^{-1}},\cdots\right)$ and
$U$ is the PMNS~\cite{Maki:1962mu} neutrino mixing matrix.  Henceforth
we will consider the case of three scalar singlets, $\alpha=1,2,3$,
where the Yukawa couplings take the form
\begin{align}
\label{eq:Y_CIww}
 h_{i\alpha}=&\frac{\sqrt{m_{\nu 1}}{\mathcal{R}}_{\alpha 1}U_{i1}^*+\sqrt{m_{\nu 2}}{\mathcal{R}}_{\alpha 2} U^{*}_{i2}+ \sqrt{m_{\nu 3}}{\mathcal{R}}_{\alpha 3} U^{*}_{i3}}{\sqrt{\Lambda_\alpha}}.
\end{align}
In the above equation, the $3\times3$ matrix
$\boldsymbol{\mathcal{R}}$ can be casted in terms of three rotation
angles $\theta_{23},\theta_{13},\theta_{12}$, which are assumed to be real. 
It is worth mentioning that for the case two scalar singlets
$\alpha=1,2$ a viable scenario is also possible with the remarks that
one massless neutrino is obtained. 
To fully exploit the generality of $h_{i\alpha}$ couplings obtained
from~\eqref{eq:Y_CIww}, we stick to the case with three scalar singlets.

In summary, the set of input parameters of the model are the scalar masses
$m_{S_\alpha}$, $M_N$, $M_D$, $\lambda$,
$\tan\beta$, the lightest neutrino mass $m_{\nu1}$, the three rotation 
angles present in $\boldsymbol{\mathcal{R}}$  
and $\lambda^{SH}_{\alpha\beta}$\footnote{The couplings $\lambda^{S}_{\alpha\beta\gamma\delta}$ are irrelevant for phenomenological purposes.}. With no lose of generality we assume for the latter to be small $\lambda^{SH}_{\alpha\beta}\lesssim0.01$, except for the case of scalar dark matter where $\lambda^{SH}_{11}$ is set to give 
the proper relic density.

\begin{figure}
\centering
(a) \hfill (b)\\
\includegraphics[scale=0.58]{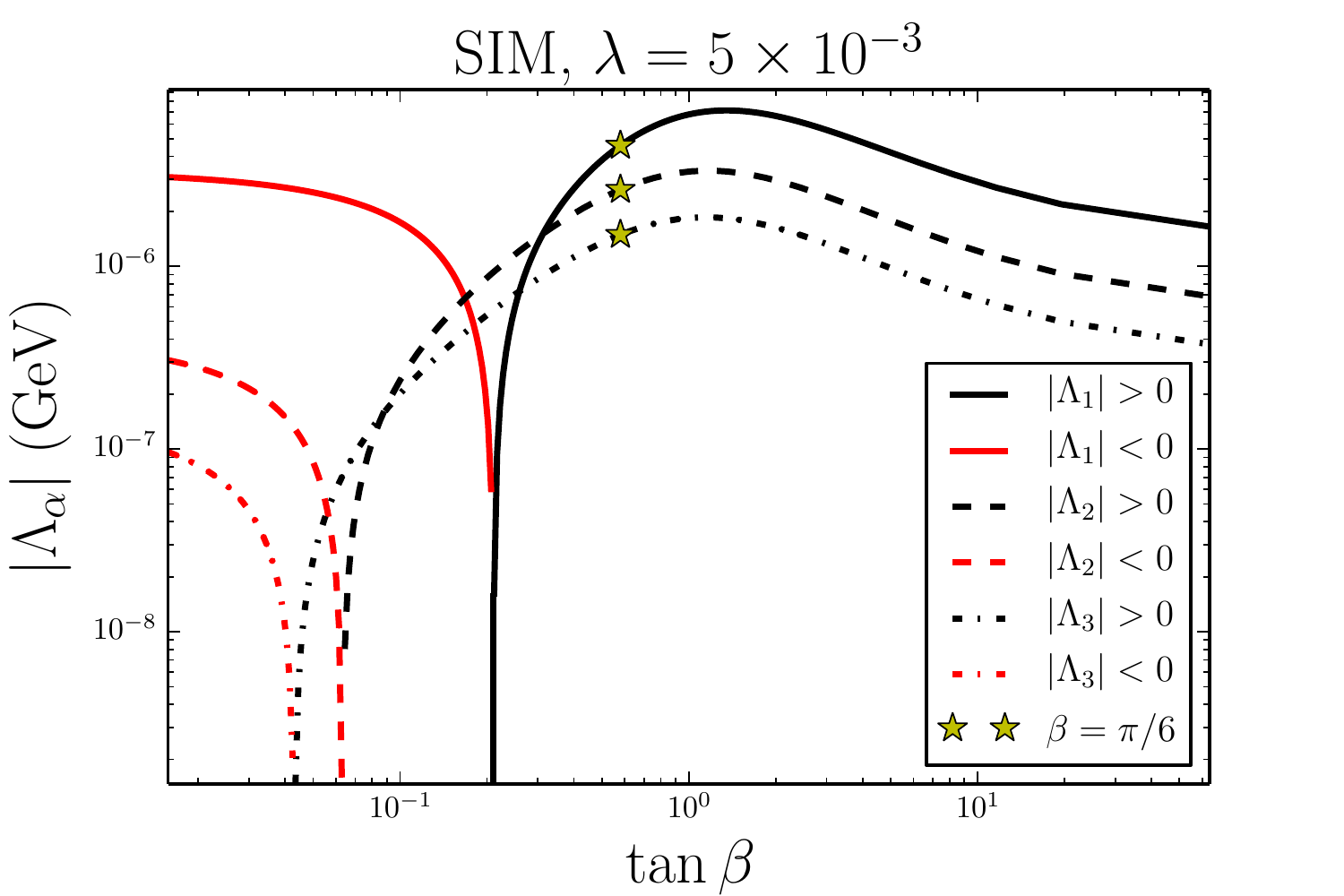}    
\includegraphics[scale=0.58]{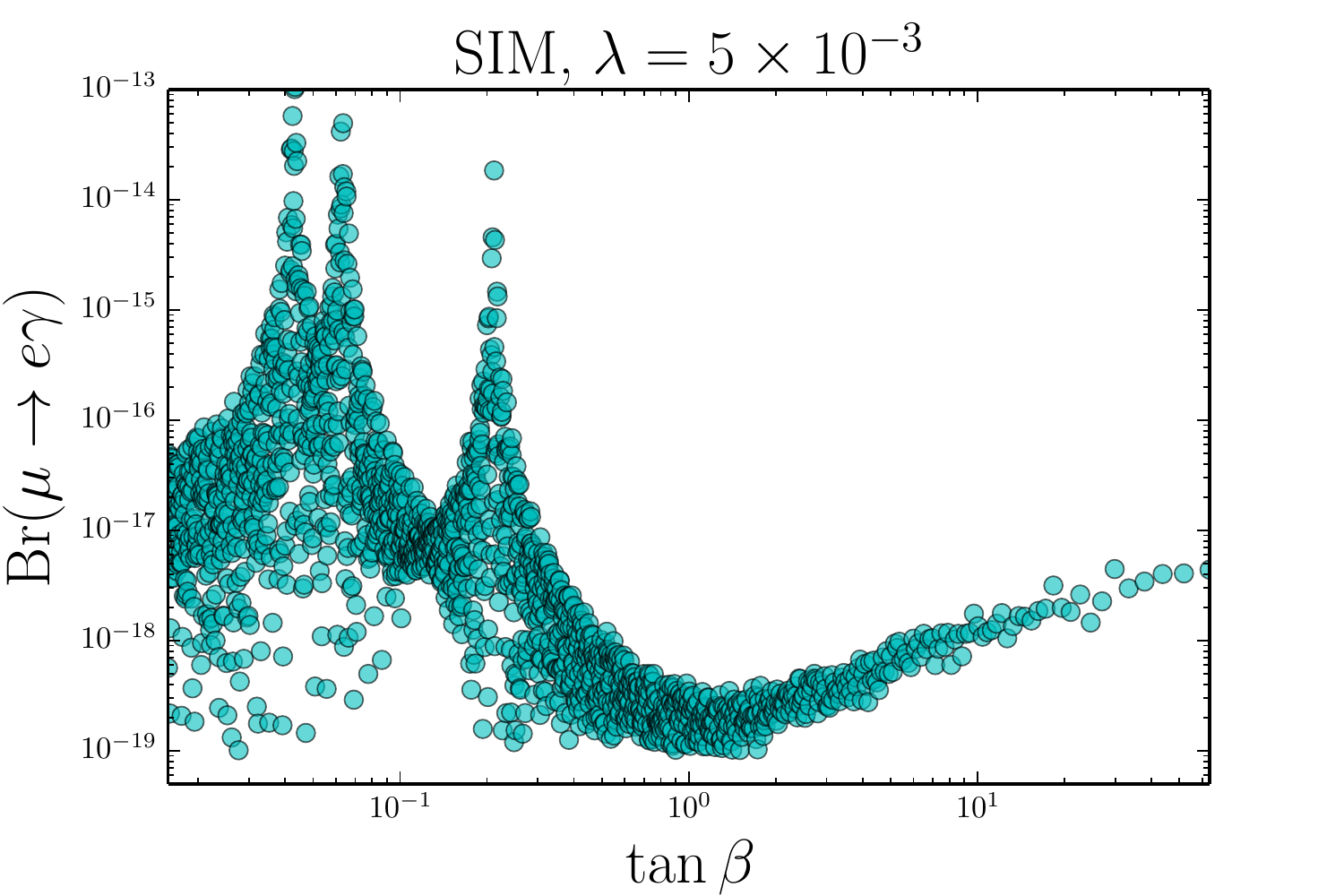}
\caption{$\tan\beta$ dependence of (a) $\Lambda_a$ and (b) $\operatorname{Br}(\mu \rightarrow e \gamma)$, for the set of input masses in eq.~\eqref{eq:bp} with
$\lambda=\num{5E-3}$.}
\label{fig:Lambdaa}
\end{figure}

In order to have an approximate expression for $\Lambda_\alpha$ in terms of this set of input parameters, we can use the identity~(\ref{eq:divcan}) to obtain
\begin{align*}
\Lambda_\alpha=&\frac{1}{16\pi^2}\left\{N_{31}^2\,m^\chi_1\left[ f(m_{S_\alpha},m^\chi_1)-f(m_{S_\alpha},m^\chi_3)\right]
                                +N_{32}^2\,m^\chi_2\left[f(m_{S_\alpha},m^\chi_2)-f(m_{S_\alpha},m^\chi_3)\right]\right\}\,.
\end{align*}
The expression for the matrix elements $N_{31}^2$ at $\mathcal{O}\left( m_{\lambda}^2 \right)$ are given in the Appendix \ref{sec:analyt-form-mass}. Since $N_{31}^2$ and $f(m_{S_\alpha},m^\chi_2)-f(m_{S_\alpha},m^\chi_3)$ are already $\mathcal{O}\left( m_{\lambda}^2 \right)$, we can use the leading order values for the other masses and mixings parameters to obtain
\begin{align*}
  \Lambda_\alpha\approx&\frac{1}{16\pi^2}\left\{N_{31}^2 M_N\left[ f(m_{S_\alpha},M_N)-f(m_{S_\alpha},M_D)\right]
                                +\frac{1}{2}M_D\left[f(m_{S_\alpha},m^\chi_2)-f(m_{S_\alpha},m^\chi_3)\right]\right\}
+\mathcal{O}\left( m_\lambda^4 \right)\,.
\end{align*}
With the last two approximate formulas for masses in \eqref{eq:ml2},  and the $N_{31}^2$  mixing in~\eqref{eq:mixl2}, we have
\begin{align}
\label{eq:lambdaappr}
16\pi^2\frac{\Lambda_\alpha}{m_{\lambda}^{2}}\approx & \left(\frac{M_{D} \cos\beta + M_{N} \sin\beta}{M_{D}^{2} - M_{N}^{2}}\right)^{2}
                                  M_{N}\left[f(m_{S_\alpha},M_N)-f(m_{S_\alpha},M_D)\right]\nonumber\\
&+\frac{ M_{D}^{2}\left[M_{D} \sin\left(2 \beta \right ) + M_{N}\right] 
        }{\left(M_{D}^{2} - M_{N}^{2}\right) \left(M_{D}^{2} - m_{S_\alpha}^{2}\right)^{2}} \left\{M_{D}^{2} - m_{S_\alpha}^{2} \left[\log{\left (\frac{M_{D}^{2}}{m_{S_\alpha}^{2}} \right )} + 1\right]\right\}+\mathcal{O}\left( m_\lambda^2 \right)\,.
\end{align}
To illustrate the dependence in $\tan\beta$ of $\Lambda_{\alpha}$, we consider the following set of input masses (SIM) compatible with singlet scalar dark matter:
\begin{align}
\label{eq:bp}
  m_{S_1}&=\SI{60}{GeV} & m_{S_2}=&\SI{800}{GeV} & m_{S_3}=&\SI{1500}{GeV}\, \nonumber\\
  m_N&=\SI{100}{GeV} & m_D=&\SI{550}{GeV}\,.
\end{align}
The results for $\lambda=\num{5E-3}$ are shown in
figure~\ref{fig:Lambdaa}(a). For large values of $\tan\beta$, the
$\Lambda_{\alpha}$ are positive. However, there are specific values of $\tan\beta$ for which each $\Lambda_{\alpha}$ 
goes to zero and turn to negative values as illustrated by the red lines in the plot. 
The specific point with $\beta=\pi/6$ is illustrated by the yellow stars in the figure.

\section{Lepton flavor violation}
\label{sec:lept-flav-viol}
The size of the lepton flavor violation (LFV) is controlled by the lepton
number violating couplings $h_{i\alpha}$.  
From the approximate expression for $\Lambda_{\alpha}$
in~\eqref{eq:lambdaappr} and the analysis of the previous section, we
will show that these couplings are inversely related to the Yukawa
coupling strength $\lambda$.
Since in SDFDM the observed dark matter abundance is typically
obtained for $\lambda\gtrsim 0.1$~\cite{Cheung:2013dua}, the lepton
flavor observables are not expected to give better constraints than
the obtained from direct detection experiments. Therefore we will
focus our discussion of LFV in regions of the parameter space
where $S_1$ is the dark matter candidate.

It is well known LFV processes put severe constraints on the LFV
couplings and in general on the model's parameter space. 
One of the most restrictive LFV processes is the radiative muon decay
$\mu\to e\gamma$, which in the present model is mediated by same
particles present in the internal lines of the one-loop neutrino mass
diagram. 
The corresponding expression for the branching ratio reads
\DetailsMutoEGamma
\begin{align}
\label{eq:muegamma}
\operatorname{Br}(\mu \rightarrow e \gamma)=&\dfrac{3}{4}\dfrac{\alpha_{\text{em}}}{16 \pi G_F^2}\left|\sum_{\alpha}
h_{1\alpha}\frac{F\left(M_D^2/m_{S_{\alpha}}^2  \right) }{m_{S_\alpha}^2}h_{2\alpha}^{*}  \right|^2 ,
\end{align}
where
\begin{align}
F(x)=\dfrac{x^3-6x^2+3x+2+6x\ln x}{6(x-1)^4}\,.
\end{align}
With the implementation of the model in the
\texttt{BSM-Toolbox}~\cite{Staub:2011dp} of
\texttt{SARAH}~\cite{Staub:2008uz,Staub:2013tta}, we have crosschecked
the one-loop results for both neutrino masses and $\operatorname{Br}(\mu
\rightarrow e \gamma)$.
Moreover, with the \texttt{SARAH}
\texttt{FlavorKit}~\cite{Porod:2014xia}, we have also checked that the
most restrictive lepton flavor violating process in the scan to be
described below, is just Br($\mu\to e\gamma$).
From eq.~\eqref{eq:Mnuij}, we obtain
\begin{align}
  M^{\nu}_{12}=\sum_{\alpha} h_{1\alpha} \Lambda_{\alpha}
  h_{2\alpha}\sim \text{constant.}
\end{align}
Comparing this result with the corresponding combination of couplings
in the expression for Br$(\mu\to e\gamma)$ in eq.~\eqref{eq:muegamma},
we expect that for a set of fixed input masses $\operatorname{Br}(\mu
\rightarrow e \gamma)$ turns to be inversely proportional to
$\Lambda_\alpha^2$.
This is illustrated in figure~\ref{fig:Lambdaa}(b)
for $\lambda=\num{5E-3}$, where the scatter plot of
$\operatorname{Br}(\mu\to e \gamma)$ is shown for the same range of $\tan\beta$ values 
 than in figure~\ref{fig:Lambdaa}(a).  In such a case, once
$h_{i\alpha}$ are obtained from the Casas-Ibarra  parametrization, the
specific hierarchy of $\Lambda_{\alpha}$ fix the several contributions
to $\operatorname{Br}(\mu \rightarrow e \gamma)$.  The dispersion of
the points is due to the 3-$\sigma$ variation of neutrino oscillation
data~\cite{Forero:2014bxa} used in the numerical implementation of the Casas-Ibarra
 parametrization, along with the random variation of the parameters of
$\boldsymbol{\mathcal{R}}$. The minimum value of
$\operatorname{Br}(\mu \rightarrow e \gamma)$ around $\tan\beta=1$
corresponds to the maximum value of $\Lambda_{\alpha}$, while the
maximum values happen at the cancellation points of each
$\Lambda_{\alpha}$. In the subsequent analysis, and for a fixed SIM and $\lambda$, we allow 
for cancellations only by two orders of magnitude from the maximum
value of each $\Lambda_{\alpha}$. 

\begin{figure}
  \centering
  \includegraphics[scale=0.7]{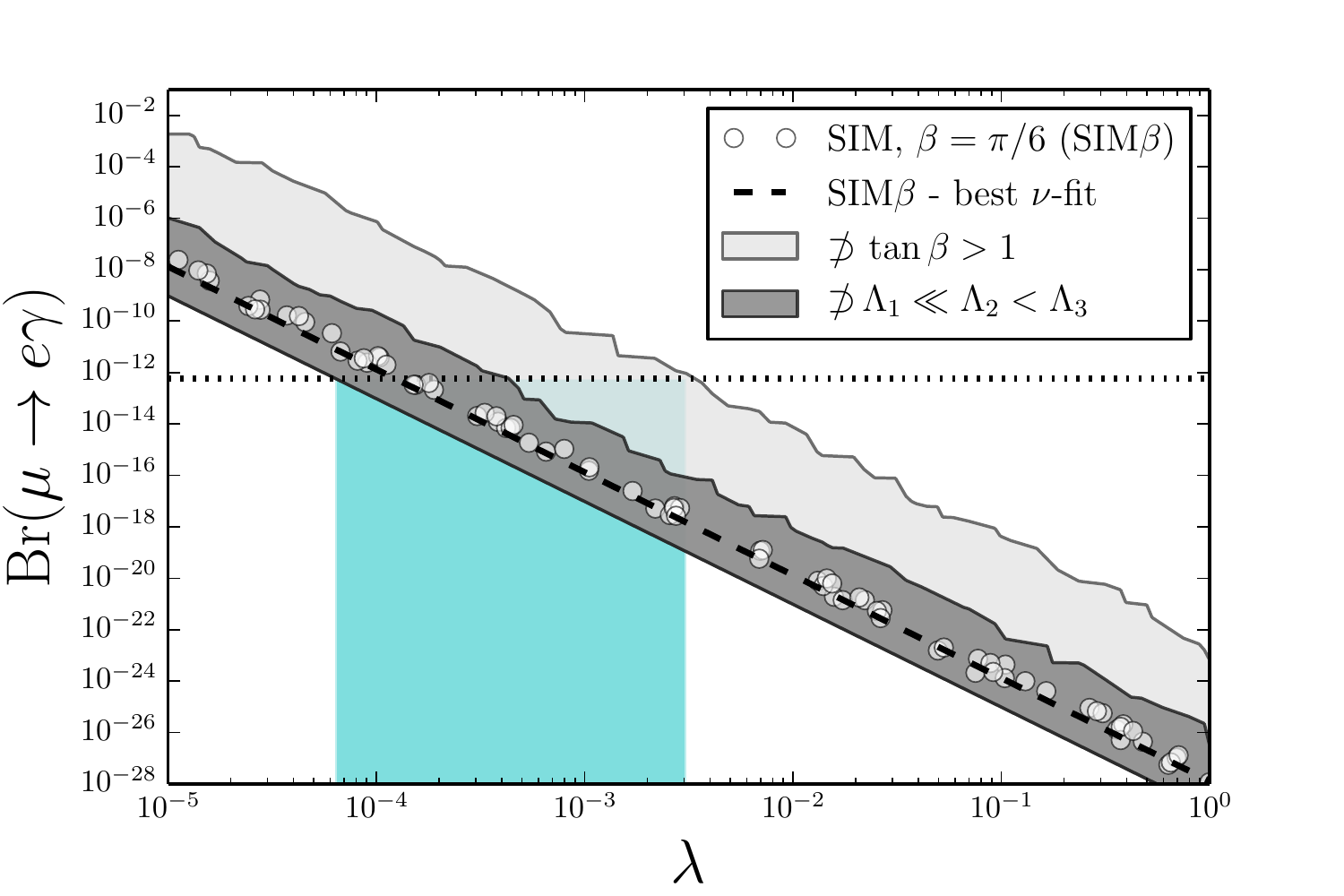}
  \caption{$\operatorname{Br}(\mu \rightarrow e \gamma)$ in terms of
    the Yukawa coupling strength $\lambda$ for the SIM in
    eq.~\eqref{eq:bp} with $\beta=\pi/6$, and the general scan
    described in the text.}
  \label{fig:brmuegamma}
\end{figure}

The full scan of the input masses up to $\SI{2}{TeV}$, with
$m_{S_1}>\SI{53}{GeV}$~\cite{Abe:2014gua} as the dark matter candidate,
$M_D>\SI{100}{GeV}$ to satisfy LEP constraints, and
$10^{-2}\le\tan\beta\le 10^2$, give to arise the dark-gray plus
light-gray regions in figure~\ref{fig:brmuegamma}. In particular, the
$\lambda$ variation for the SIM with $\beta=\pi/6$, denoted by yellow
stars in figure~\ref{fig:Lambdaa}(a), is illustrated
with the white dots in figure~\ref{fig:brmuegamma}.   The
corresponding dashed line is obtained for the best-fit values of the
neutrino oscillation data and $\boldsymbol{\mathcal{R}}$ fixed to the
identity. The horizontal dotted line in the plot corresponds to the
current experimental bound for $\operatorname{Br}(\mu\to e
\gamma)<\num{5.7E-13}$ at $90\%\,\si{CL}$~\cite{Adam:2013mnn}.  
The upper part of the 
light-gray region is restricted by our imposition to avoid too strong
cancellation in  $\Lambda_{\alpha}$ . We check that for all the sets of
input masses in the random scan, this cancellation region always
happens when $\tan\beta<1$. In this way, points with $\tan\beta>1$ are
absent from the light-gray region, as labeled in
figure~\ref{fig:brmuegamma}. For the same reason, in the dark-gray
region there are not points with $\Lambda_{\alpha}\ll
\Lambda_{\beta}\sim \Lambda_\gamma$ ($\alpha\ne\beta\ne\gamma$). We can check for example that
points with $\Lambda_1\ll \Lambda_2<\Lambda_3$ are absent inside the
dark-gray region of figure~\ref{fig:brmuegamma}.

The lower part of the dark-gray region is saturated by the values of
$M_{D}=\SI{2}{TeV}$, and gives rise to the lower bound $\lambda\gtrsim
\num{6E-5}$. 
With our restriction in the cancellation of
$\Lambda_{\alpha}$,  points in the scan with $\lambda\lesssim\num{3E-3}$ 
can be excluded from the Br$(\mu\to e\gamma)$ limit.

\section{Collider phenomenology}
\label{sec:collider-phenomenology}

The LHC phenomenology in the case of the singlet-doublet fermion
dark matter was already analyzed in~\cite{Abe:2014gua}. Their conclusion, is that
the recast of the current LHC data is easier to evade, but the
long-rung prospects are promising, since the region $M_N,m_\lambda\ll M_D$ could be 
probed up to $M_D\lesssim 600-\SI{700}{GeV}$ for the 14-TeV run of the LHC with 
$\SI{3000}{fb}^{-1}$. 

\begin{figure}[h]
\begin{center}
\includegraphics[scale=0.4]{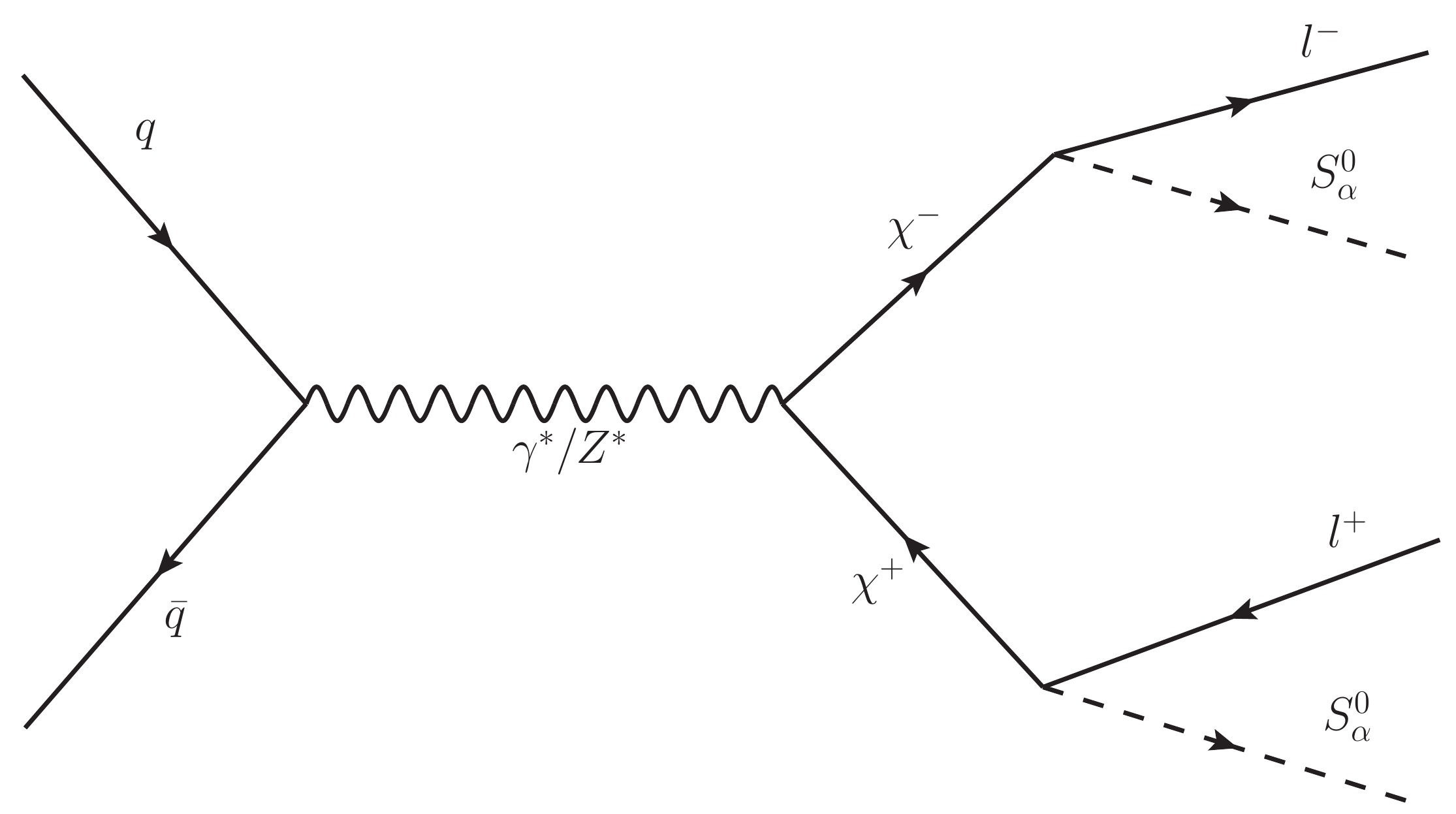}
\caption{Feynman diagram for $pp \rightarrow \chi^+\chi^- \rightarrow l^+l^- S_{\alpha}^0S_{\alpha}^0$ .}
\label{fig:chain_signal}
\end{center}
\end{figure}

On the other hand, in the case of the singlet scalar dark matter, the
main production processes associated with the new fermions remain the
same, but there are new signals from the mediation, or presence in the
final decay chains, of the new scalars.
The most promising possibility is the dilepton plus missing
transverse energy signal coming from the production of
charged fermions decaying into leptons and the lightest scalar.
This signal can be important when $\lambda$ is
not too large, $\lambda\lesssim 0.1$, and $M_N\gtrsim M_D$. 
For a fixed set of input parameters, the random phases in the
Casas-Ibarra can be chosen to have all the possibilities in the lepton
flavor space associated with the coupling $h_{i1}$, with
$i=e,\mu,\tau$. 
In view of that, we will focus in the best scenario where
$\operatorname{Br}(\chi^\pm\to e^\pm \, S_1 )\approx 1$.  
The Feynman diagram for the processes is displayed in
figure~\ref{fig:chain_signal}.

The mass of the  charged  Dirac fermion $\chi^{\pm}$, can be
constrained from dilepton plus missing transverse energy  searches at the  LHC.
In~\cite{Aad:2014vma}, this kind of signals was used by the ATLAS
collaboration to establish bounds on the slepton masses from the
search for $pp \rightarrow \tilde{l}^+\tilde{l}^- \rightarrow l^+l^-
\tilde{\chi}^0\tilde{\chi}^0$, where $\tilde{l}^{\pm}$ are the
sleptons, $\tilde{\chi}^{0}$ are the neutralinos and $l^-$ is $e^-$ or
$\mu^-$.  
Purely left-handed sleptons produced and decaying this way, have been
excluded up to masses of about $300$ GeV at 95\% CL, from the data
with integrated luminosity of 20.3 fb$^{-1}$ and the $pp$ collision
energy of 8 TeV. 
This corresponds to an excluded cross section of $\SI{1.4}{fb}$ at NLO
calculated with \texttt{PROSPINO}~\cite{Beenakker:1996ed}.

In the present model, the charged fermion field may decay in the
mode $\chi^{\pm} \rightarrow l_i^{\pm}S_1^0$ which are
proportional to the Yukawa couplings $h_{i1}$. 
Therefore, a similar final state as in the slepton pair production is
obtained through the process $pp \rightarrow \chi^+\chi^- \rightarrow
l^+l^- S_1^0 S_1^0$, as can be seen in figure~\ref{fig:chain_signal}.

In this case, the excluded cross section of this process can be estimated from:

\begin{align}
\label{eq:exccs}
\sigma(pp \rightarrow l^+l^-S_{\alpha}^0S_{\alpha}^0)=\sigma(pp \rightarrow \chi^+\chi^-)\times \operatorname{Br}(\chi^{\pm} \rightarrow l^{\pm}S_{\alpha}^0)^2 ,
\end{align}
where $\sigma(pp \rightarrow \chi^+\chi^-)$ is the pair production
cross section of charged Dirac fermion, and
$\operatorname{Br}(\chi^{\pm} \rightarrow l^{\pm}S_{\alpha}^0)$ is the
branching fraction for $\chi^{\pm} \rightarrow l^{\pm}S_{\alpha}^0$
mode.

\begin{figure}[h]
\begin{center}
\includegraphics[scale=0.5]{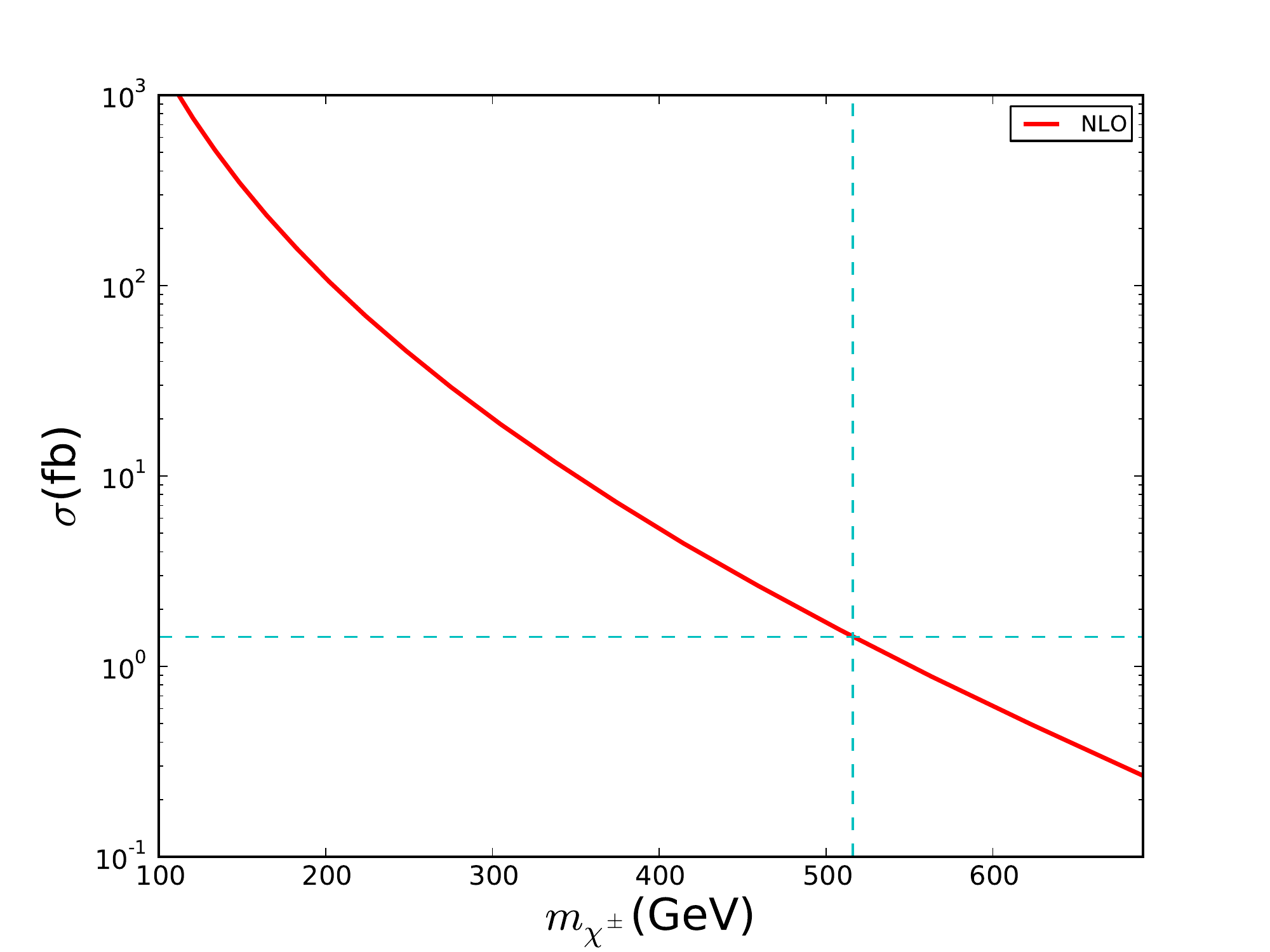}
\caption{NLO cross section for the charged Dirac fermion pair
  production at the LHC with $pp$ collisions at $\sqrt{s}=8$ TeV. 
  The horizontal dashed line for the excluded cross section of
  $\SI{1.4}{fb}$, corresponds to the mass about $\SI{510}{GeV}$
  illustrated by the vertical dashed line.}
\label{fig:chargino_production}
\end{center}
\end{figure}

The pair production of charged Dirac fermions can be
calculated in the pure-higgsino limit of the minimal supersymmetric
standard model. 
The NLO cross section calculated with \texttt{PROSPINO} is displayed
in figure \ref{fig:chargino_production} as a function of the charged
Dirac fermion.

For points in the parameter space where the Casas-Ibarra solution is
chosen such that $\operatorname{Br}(\chi^\pm\to e^\pm\,S_1)\approx 1$,
and assuming the same efficiency as for the dilepton plus missing
transverse energy signal coming from left-sleptons in
eq.~\eqref{eq:exccs}, the charged Dirac fermions of the present model
can be excluded up to $\SI{510}{GeV}$, as illustrated in
figure~\ref{fig:chargino_production}.

Note that many points in the scan of figure~\ref{fig:brmuegamma} with
$\lambda\lesssim 0.1$ and featuring $m_{S_1}\ll M_D$, could be
excluded by this LHC constraint. 
However, a detailed analysis of the restriction from the Run~I of the
LHC, in the full parameter space of the model, is beyond the scope of this
work.

\section{Singlet-doublet fermion dark matter}
\label{sec:singlet-doublet-dark}
In this model, the role of the dark matter particle can be played by
either the lightest of the fermions $\chi_{\text{LOP}}$ or the lightest of the
scalars $S_1$. 
In the latter case, the present model resembles the
singlet scalar DM model
\cite{Silveira:1985rk,McDonald:1993ex,Burgess:2000yq} as long as the
other $Z_2$-odd particles do not contribute to the total annihilation
cross section of $S_1$, namely through to the addition of new
(co)annihilation channels. 
Therefore, by choosing a non degenerate mass spectrum and small Yukawa
couplings (which is in agreement with neutrino masses) the effects of these particles on dark matter can be neglected. 
Hence we expect that the dark matter phenomenology to be similar to
that of the SSDM \cite{Cline:2013gha}. 

On the other hand, regarding the case of fermion DM, the present model includes the
singlet doublet fermion DM model
\cite{ArkaniHamed:2005yv,Mahbubani:2005pt,D'Eramo:2007ga,Enberg:2007rp,Cohen:2011ec,Cheung:2013dua}.
In such scenario, when the dark matter candidate is mainly singlet
(doublet) the relic density is in general rather large (small). 
In particular, a pure doublet has the proper relic density for
$M_{D}\sim\SI{1}{TeV}$~\cite{Mahbubani:2005pt,Cheung:2013dua,Chattopadhyay:2005mv}
with decreasing  values as $M_D$ decreases.  
Nonetheless, in the present model we have the additional possibility
of coannihilations between the $Z_2$-odd scalars and fermions. 
In this work, we explore at what extent coannihilation with
scalars may allow to recover pure-doublet DM regions with
$M_D\lesssim\SI{1}{TeV}$ and $\lambda\lesssim0.3$, while keeping the
proper relic density.  
Hereafter, we focus in that specific region.

In the simple radiative seesaw model with inert
doublet scalar dark matter, the coannihilations with singlet fermions can enhance
rather than reduce the relic density, as shown in~\cite{Klasen:2013jpa}. That work also presented a review of the several
models~\cite{Servant:2002aq,Kong:2005hn,Burnell:2005hm,Edsjo:2003us,Profumo:2006bx}
where such an enhancement also occurs.
In particular, supersymmetric models where
the neutralino is higgsino-like were considered in~\cite{Profumo:2006bx} and it
was shown that slepton coannihilations not only lead to an increase in
the relic density but also to an enhancement in the predicted
indirect detection signals. 
Below, we show that the singlet scalars can play the role of the
sleptons in our generalization of the higgsino-like dark matter with
radiative neutrino masses.

\begin{figure}
  \centering
  \includegraphics[scale=0.5]{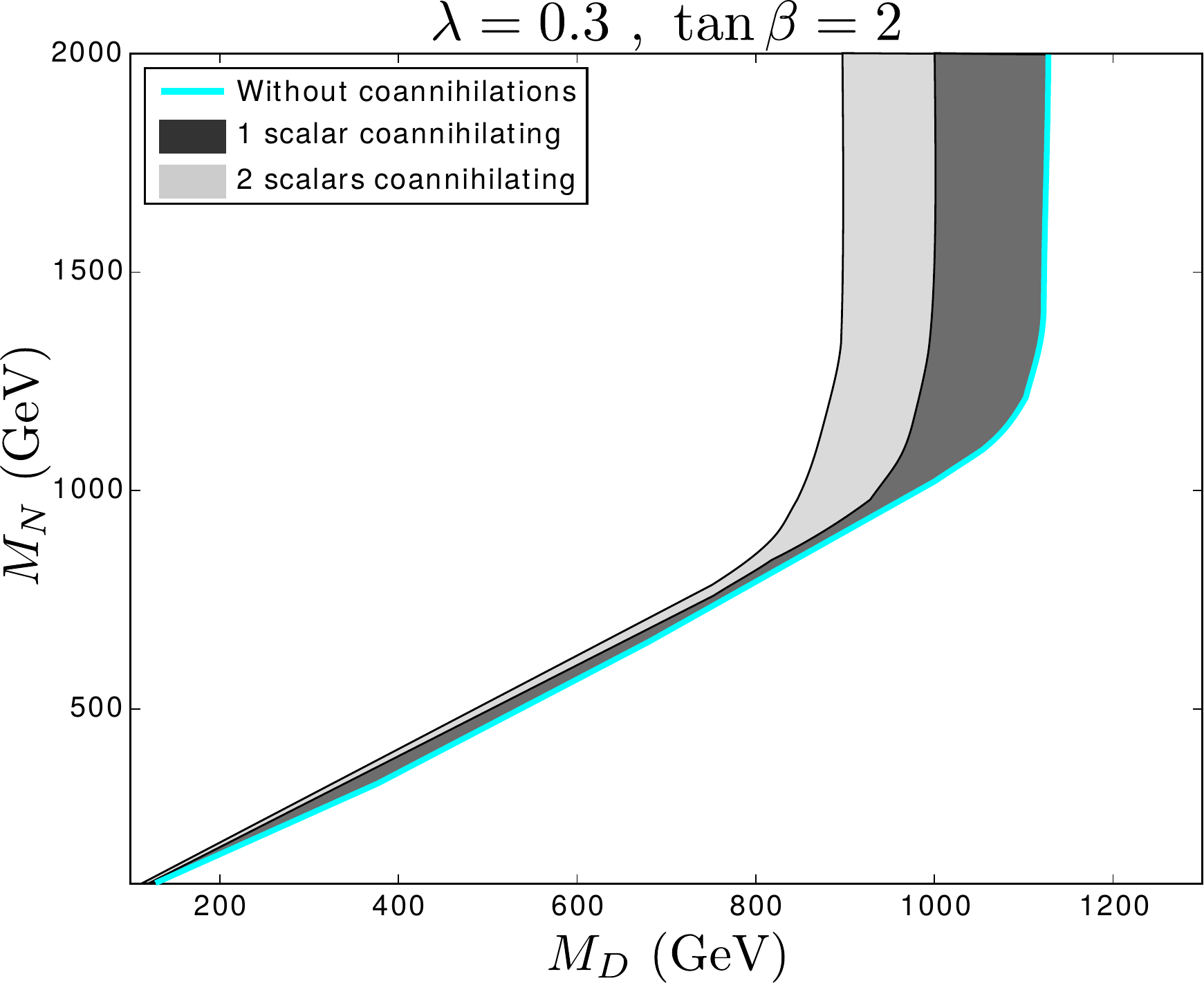}
  \caption{Regions consistent with the observed relic density for $\lambda=0.3$
   and $\tan\beta=2$.    The solid cyan line corresponds to the observed relic density
   without coannihilations which was shown to be compatible with the
   current direct detection bounds from LUX~\cite{Akerib:2013tjd}
   in~\cite{Cheung:2013dua}.
   The effect of the coannihilations with the new scalars is shown for
   a mass degeneracy of 0.1 to 10\% between the scalars and the DM
   candidate. 
   The dark-gray region corresponds to coannihilations with one scalar
   singlet, while the dark plus light-gray regions correspond to
   coannihilations with two scalar singlets.  }
  \label{fig:5b}
\end{figure}

The interactions of the scalars $S_\alpha$ are described by the
$h_{i\alpha}$, $\lambda_{\alpha\beta}^{SH}$ terms in
eq.~(\ref{eq:lt13a}). 
It turns out that Yukawa interactions are suppressed by neutrino
masses ($h_{i\alpha}\lesssim10^{-4}$) and the same occurs for the interaction
with the Higgs boson if we impose
$\lambda_{\alpha\beta}^{SH}\lesssim10^{-2}$.
In this way the coannihilating scalars $S_\alpha$ act as as parasite
degrees of freedom at freeze-out leading to an increase of the
singlet-doublet fermion relic density. 

By following the discussion in ~\cite{Klasen:2013jpa}, the maximum
enhancement of the relic density is achieved when
$\Delta_{S_\alpha}=(m_{S_{\alpha}}-m_{\text{LOP}}^{\chi})/m_{\text{LOP}}^{\chi}$
becomes negligible. 
Accordingly one can write
\begin{align}
\frac{  \Omega^{S_{\alpha}}}{\Omega^{0}}\approx\left( \frac{g_0+g_{S_{\alpha}}}{g_0} \right)^2,
\end{align}
where $\Omega^{S_{\alpha}}$ ($\Omega^{0}$) denotes the relic density
with (without) including $S_{\alpha}$ coannihilations,
$g_{S_{\alpha}}$ represents the total number of internal degrees of
freedom related to the scalars participating in the in the
coannihilation process and $g_{0}$ is the total number of internal
degrees of freedom when $\Delta_{S_\alpha}\gg1$. 
When the DM particle is pure doublet ($M_D\sim1$ TeV and $M_N\gg M_D$)
the fermion masses are $m_1^{\chi}=M_N,\,m_{2,3}^{\chi}\approx
m_{\chi^\pm}=M_D$ and therefore
$g_{0}=g_{\chi_2}+g_{\chi_3}+g_{\chi^\pm}=8$.
Since each real scalar have one degree of freedom we have
$g_{S_{\alpha}}=1,2,3$ depending on the number of scalars
coannihilating from which it follows that the maximum enhancement is
$\Omega^{S_{\alpha}}/\Omega^{0}=1.27,\, 1.56,\,1.89$, respectively. 
This enhancement results in that for the present model with
doublet-like DM and $\lambda\lesssim0.3$ the $M_D$ required to explain
the correct relic density lies in the range $[0.9,1.1]$~TeV instead of
taking a single value as in the SDFDM model.  
The values inside this range, arise due to a no mass degeneracy
between the fermions and scalars. 
In figure~\ref{fig:5b} we show the effect of coannihilations on the
relic density~\footnote{The relic density is calculated with the
  \texttt{BSM-Toolbox} chain: 
\texttt{SPheno}~3.3.6~\cite{Porod:2011nf}-\texttt{MicrOMEGAs}~4.1.7~\cite{Belanger:2006is,Belanger:2014vza}.}
of $m_{\text{LOP}}^{\chi}$ for a mass degeneracy of 0.1
to 10\% between  scalar singlets and the DM candidate and for
$\lambda=0.3$ and $\tan\beta=2$. 
In particular, in the light-gray region we plot the coannihilations
with two scalars to facilitate the comparison with the results
in~\cite{Profumo:2006bx} for higgsino-like dark matter coannihilating
with a right-handed stau ($g\approx 2$ in their plots).
As expected, the upper limit in the LOP mass is about $20\%$ smaller
with respect to the case without coannihilation, and we could 
expect similar enhancements for indirect DM searches
as  in ~\cite{Profumo:2006bx} for $g\approx 2$.
Note that the impact of the $S_\alpha$ coannihilations
when $M_D,\,M_N<\SI{1}{TeV}$,  is reduced because in such case the
dark matter particle is a mixture of singlet and doublet
(well-tempered DM \cite{ArkaniHamed:2006mb}), and the non-negligible
splitting among the fermion particles $\chi$ leads to a non-zero
Boltzmann suppression. 
We have checked that the same results are obtained when
$\lambda\lesssim0.3$.

With regard to DM direct detection in the pure-doublet DM scenario
discussed above, it is not restricted by the current LUX~\cite{Akerib:2013tjd} bounds as
long as $\tan\beta>0$.  
This is due to the existence of zones, known as blind spots, where the
spin independent cross section vanishes identically and they occur
only for positive values of $\tan\beta$
\cite{Cheung:2013dua}\footnote{Note that $\tan\beta>0$ corresponds to
  $\tan\theta<0$ in notation of  \cite{Cheung:2013dua}.}. 
In consequence, the recovered pure-doublet DM regions are still viable
in light of the present results of direct searches of dark matter.  

\section{Conclusions}
\label{sec:conclusions}

We have combined the singlet-doublet fermion dark matter (SDFDM) and
the singlet scalar dark matter (SSDM) models into a framework that generates radiative neutrino masses. 
The required lepton number violation only happens if the scalars are
real.   
We have then explored the novel features of the final model in flavor
physics, collider searches, and dark matter related experiments.  
In the case of SSDM, for example, the singlet-doublet fermion mixing
cannot be too small in order to be compatible with lepton flavor
violating (LFV) observables like $\operatorname{Br}(\mu\to e\gamma)$,
while in the case of fermion dark matter the LFV constraints are
automatically satisfied.
The presence of new decay channels for the next to lightest odd
particle opens the possibility of new signals at the LHC.
In particular, when the singlet scalar is the lightest
odd-particle and the singlet-like Majorana fermion is heavier than the
charged Dirac fermion, the production of the later yields dilepton plus missing transverse energy signals. For large enough
$e^\pm$ or $\mu^\pm$ branchings, these signals could exclude charged
Dirac fermion masses
of order $\SI{500}{GeV}$ in the Run I of the LHC. 
Finally, the effect of coannihilations with the scalar singlets was
studied in the case of doublet-like fermion dark matter.  In that
case, it is possible to obtain the observed dark matter relic density
with lower values of the LOP mass.

\section{Acknowledgments}
We are very gratefully to Camilo García Cely, Enrico Nardi, Federico
von der Pahlen and specially to Carlos Yaguna for their illuminating
discussions.
DR and OZ have been partially supported by UdeA through the grants
Sostenibilidad-GFIF, CODI-2014-361 and CODI-IN650CE, and COLCIENCIAS
through the grants numbers 111-556-934918 and 111-565-842691. 
WT has been supported by grants from ISF (1989/14), US-Israel BSF
(2012383) and GIF (I-244-303.7-2013).

\appendix

\section{Analytic formulas for masses and mixing matrix of neutral fermions}
\label{sec:analyt-form-mass}

The characteristic equation of the mass matrix~\eqref{eq:Mchi}
is~\cite{Cheung:2013dua}\footnote{The analytic formulas for the neutralino
  masses and the neutralino mixing matrix was
  analyzed in \cite{ElKheishen:1992yv}.}:
\begin{details}
\begin{align}
\left[\left({M}^{\chi}_{\text{diag}}\right)_{ii}\right]^3-M_N\left[\left({M}^{\chi}_{\text{diag}}\right)_{ii}\right]^2
 -\left(M_D^2+m_{\lambda}^2\right)\left({M}^{\chi}_{\text{diag}}\right)_{ii}+\left[M_NM_D^2-m_{\lambda}^2\sin(2\beta)
   M_D\right]=0\,,
\end{align}
or
\end{details}
\begin{align*}
\left[\left({M}^{\chi}_{\text{diag}}\right)_{ii}^2-M_D^2\right]
\left[M_N-\left({M}^{\chi}_{\text{diag}}\right)_{ii}^{\phantom{2}}
\right]
+\tfrac{1}{2}m_{\lambda}^2\left[\left({M}^{\chi}_{\text{diag}}\right)_{ii}+M_D\sin 2\beta
\right]=0\,.
\end{align*}
The solutions to the cubic equation in  $\left({M}^{\chi}_{\text{diag}}\right)_{ii}$ are:
\begin{align}
m_1^\chi=&z_2+\dfrac{M_N}{3}\,,&
m_2^\chi=&z_1+\dfrac{M_N}{3}\,, &
m_3^\chi=&z_3+\dfrac{M_N}{3}\,.
\end{align}
where
\begin{align}
z_1&=\left(-\dfrac{q}{2}+\sqrt{\frac{q^2}{4}+\frac{p^3}{27}}\right)^{1/3} + \left(-\dfrac{q}{2}-\sqrt{\dfrac{q^2}{4}+\dfrac{p^3}{27}}\right)^{1/3}\nonumber\\ 
z_2&=-\frac{z_1}{2}+\sqrt{\frac{z_1^2}{4}+\frac{q}{z_1}} \nonumber\\ 
z_3&=-\frac{z_1}{2}-\sqrt{\frac{z_1^2}{4}+\frac{q}{z_1}}\nonumber\\ 
p&=-\frac{1}{3}M_N^2-\left(M_D^2+m_{\lambda}^2\right)  \nonumber\\ 
q&=-\frac{2}{27}M_N^3-\frac{1}{3}M_N\left(M_D^2+m_{\lambda}^2\right)+\left[M_NM_D^2-m_{\lambda}^2\sin(2\beta) M_D\right].
\end{align}

Notice that ${q^2}/{4}+{p^3}/{27} < 0$ and therefore, we have three real masses $m_i^\chi$ $(i=1,2,3)$.

Expanding the eigensystem in eq.~\eqref{eq:chidiag} by  assuming  that ${N}_{1i}\neq 0$, we have 

\begin{align*}
{M}^{\chi}_{21}\frac{{N}_{2i}}{{N}_{1i}}+{M}^{\chi}_{31}\frac{{N}_{3i}}{{N}_{1i}}&=-({M}^{\chi}_{11}-m_i^\chi)\nonumber \\
({M}^{\chi}_{22}-m_i^\chi)\frac{{N}_{2i}}{{N}_{1i}}+{M}^{\chi}_{32}\frac{{N}_{3i}}{{N}_{1i}}&=-{M}^{\chi}_{12} \nonumber\\
{M}^{\chi}_{23}\frac{{N}_{2i}}{{N}_{1i}}+({M}^{\chi}_{33}-m_i^\chi)\frac{{N}_{3i}}{{N}_{1i}}&=-{M}^{\chi}_{13}\,,
\end{align*}
where
\begin{align}
\label{eq:N1i}
{N}_{1i}=\left[1+\left(\frac{{N}_{2i}}{{N}_{1i}}\right)^2+\left(\frac{{N}_{3i}}{{N}_{1i}}\right)^2\right]^{-1/2}.
\end{align}

Using the matrix $\textbf{M}^{\chi}$ given in the eq. \eqref{eq:Mchi}, we get the ratios

\begin{align}
\label{eq:exNmn}
\frac{{N}_{2i}}{{N}_{1i}}
&=-\frac{m_{\lambda} \cos\beta}{m_i^\chi}+\frac{M_D}{m_i^\chi}\frac{[m_i^\chi(M_N-m_i^\chi)+m_{\lambda}^2\cos\beta^2 ]}{m_{\lambda}(m_i^\chi\sin\beta +M_D\cos\beta )}\,, \nonumber\\  
\frac{{N}_{3i}}{{N}_{1i}}
&=-\frac{[m_i^\chi(M_N-m_i^\chi)+m_{\lambda}^2 \cos\beta^2]}{m_{\lambda} (m_i^\chi\sin\beta+M_D\cos\beta)}.
\end{align}

\subsection{Approximate mixing matrix}
By using the analytical expressions for the mixing ratios of
eq.~\eqref{eq:exNmn} with the approximate eigenvalues~\eqref{eq:ml2}
in eq.~\eqref{eq:N1i}, we obtain

\begin{align}
N_{11}^2=& 1-\frac{\left[M_D^2+M_N^2+2M_DM_N\sin(2\beta)\right]m^2_{\lambda}}{(M_D^2-M_N^2)^2}+\mathcal{O}\left( m_{\lambda}^4 \right)\nonumber\\
N_{12}^2 =&   \frac{[\sin (2 \beta )+1] m_{\lambda }^2}{2 \left(M_N-M_D\right)^2}+\mathcal{O}\left( m_{\lambda}^4 \right)\nonumber\\
N_{13}^2=&  -\frac{[\sin (2 \beta )-1] m_{\lambda }^2}{2 \left(M_D+M_N\right)^2}+\mathcal{O}\left( m_{\lambda}^4 \right)\,.
\end{align}
\begin{align}
N_{21}^2=&\frac{m_{\lambda }^2 \left(\sin\beta  M_D+\cos\beta
   M_N\right)^2}{\left(M_N^2-M_D^2\right)^2}+\mathcal{O}\left( m_{\lambda}^4 \right)\nonumber\\
N_{22}^2=&\frac{1}{2}-\frac{m_{\lambda }^2 (\sin\beta+\cos\beta) \left[\cos\beta M_N-\sin\beta  \left(M_N-2 M_D\right)\right]}{4 M_D \left(M_N-M_D\right)^2}+\mathcal{O}\left( m_{\lambda}^4 \right)\nonumber\\
   N_{23}^2=&\frac{1}{2}+\frac{m_{\lambda }^2 (\cos\beta-\sin\beta) \left[\sin\beta \left(2
   M_D+M_N\right)+\cos\beta M_N\right]}{4 M_D \left(M_D+M_N\right)^2}+\mathcal{O}\left( m_{\lambda}^4 \right).
\end{align}

\begin{align}
\label{eq:mixl2}
N_{31}^2=&\left(\frac{M_{D} \cos\beta + M_{N} \sin\beta}{M_{N}^{2}- M_{D}^{2}} \right)^{2}   m_{\lambda}^{2}
+\mathcal{O}\left( m_{\lambda}^4 \right)\nonumber\\
N_{32}^2=&\frac{1}{2}-\frac{\left[M_{N} \sin\beta - \left( M_{N}- 2 M_{D}\right) \cos\beta\right] \left(\cos\beta+\sin\beta \right)}{4 M_{D} \left(M_{N}- M_{D}\right)^{2}}\,m_{\lambda}^{2}+\mathcal{O}\left( m_{\lambda}^4 \right)  \nonumber\\
N_{33}^2=&\frac{1}{2}- \frac{\left[M_{N} \sin\beta + \left(M_{N} + 2 M_{D}\right) \cos\beta\right] \left(\cos\beta- \sin\beta \right)}{4 M_{D} \left(M_{N} + M_{D} \right)^{2}} \,m_{\lambda}^{2}+\mathcal{O}\left( m_{\lambda}^4 \right).
\end{align}
In particular, with eq.~\eqref{eq:ml2} and the expressions for $N_{3i}^2$, the identity \eqref{eq:divcan}
is satisfied up to terms of order $\mathcal{O}\left( m_{\lambda}^4
\right)$.

\bibliographystyle{h-physrev4}%apsrev4-1long
\bibliography{susy}

\end{document}